\documentclass[useAMS, usenatbib]{mn2e}
\usepackage{graphicx}
\usepackage{journals}

\def\l{~$\lambda$}
\def\ll{~$\lambda\lambda$}

\def\hbet{H$\beta$}

\def\hdelta{H$\delta$}
\def\hgam{H$\gamma$}

\def\hea{He\,{\sc i}}
\def\heb{He\,{\sc ii}}
\def\nc{N\,{\sc iii}}
\def\nd{N\,{\sc iv}}
\def\ne{N\,{\sc v}}

\def\sid{Si\,{\sc iv}}

\def\kms{km\,s$^{-1}$}

\def\msun{M$_{\odot}$}

\def\gtrsim{\mathrel{\hbox{\rlap{\hbox{\lower3pt\hbox{$\sim$}}}\hbox{\raise2pt\hbox{$>$}}}}}
\def\lesssim{\mathrel{\hbox{\rlap{\hbox{\lower3pt\hbox{$\sim$}}}\hbox{\raise2pt\hbox{$<$}}}}}
\def\wr21a{WR21a}

\title[The mass of WR21a]{The mass of the very massive binary WR21a\thanks{Based on observations collected under program IDs 085.D-0704, 086.D-0446, 089.C-0874 and 090.D-0212 at the European Organisation for Astronomical Research in the Southern Hemisphere, Chile.}}
\author[F. Tramper et al.]{F. Tramper$^{1}$,
H. Sana$^{2}$,
N.E. Fitzsimons$^1$,
A. de Koter$^{1,2}$,
L. Kaper$^1$,
\newauthor
L. Mahy$^3$,
A. Moffat$^4$ \\
$^{1}$ Anton Pannenkoek Institute for Astronomy, University of Amsterdam, 1090 GE Amsterdam, The Netherlands \\
$^{2}$ Instituut voor Sterrenkunde, Universiteit Leuven, Celestijnenlaan 200 D, B-3001, Leuven, Belgium \\
$^{3}$ Institut d'Astrophysique, Universite de Li\`ege, All\'ee du 6 Ao\^ut 17, B-4000 Li\`ege, Belgium \\
$^{4}$ D\'epartment de Physique, Universit\'e de Montr\'eal and Centre de Recherche en Astrophysique du Qu\'ebec, C. P. 6128, succ. centre-ville, Montr\'eal (Qc) H3C 3J7, Canada}

\begin{document}

\date{---. ---; ---}

\pagerange{\pageref{firstpage}--\pageref{lastpage}} \pubyear{2002}

\maketitle

\label{firstpage}

\begin{abstract}
We present multi-epoch spectroscopic observations of the massive binary system WR21a, which include the January 2011 periastron passage. Our spectra reveal multiple SB2 lines and facilitate an accurate determination of the orbit and the spectral types of the components. We obtain minimum masses of $64.4\pm4.8$ \msun \ and $36.3\pm1.7$ \msun \ for the two components of WR21a. Using disentangled spectra of the individual components, we derive spectral types of O3/WN5ha and O3Vz~((f*)) for the primary and secondary, respectively. Using the spectral type of the secondary as an indication for its mass, we estimate an orbital inclination of $i=58.8\pm2.5$\degr\ and absolute masses of $103.6\pm10.2$ \msun \ and $58.3\pm3.7$ \msun, in agreement with the luminosity of the system. The spectral types of the WR21a components indicate that the stars are very young (1$-$2 Myr), similar to the age of the nearby Westerlund 2 cluster. We use evolutionary tracks to determine the mass-luminosity relation for the total system mass. We find that for a distance of 8 kpc and an age of 1.5 Myr, the derived absolute masses are in good agreement with those from evolutionary predictions.

\end{abstract}


\section{Introduction}

\begin{figure*}
\centering
\includegraphics[width=0.85\textwidth]{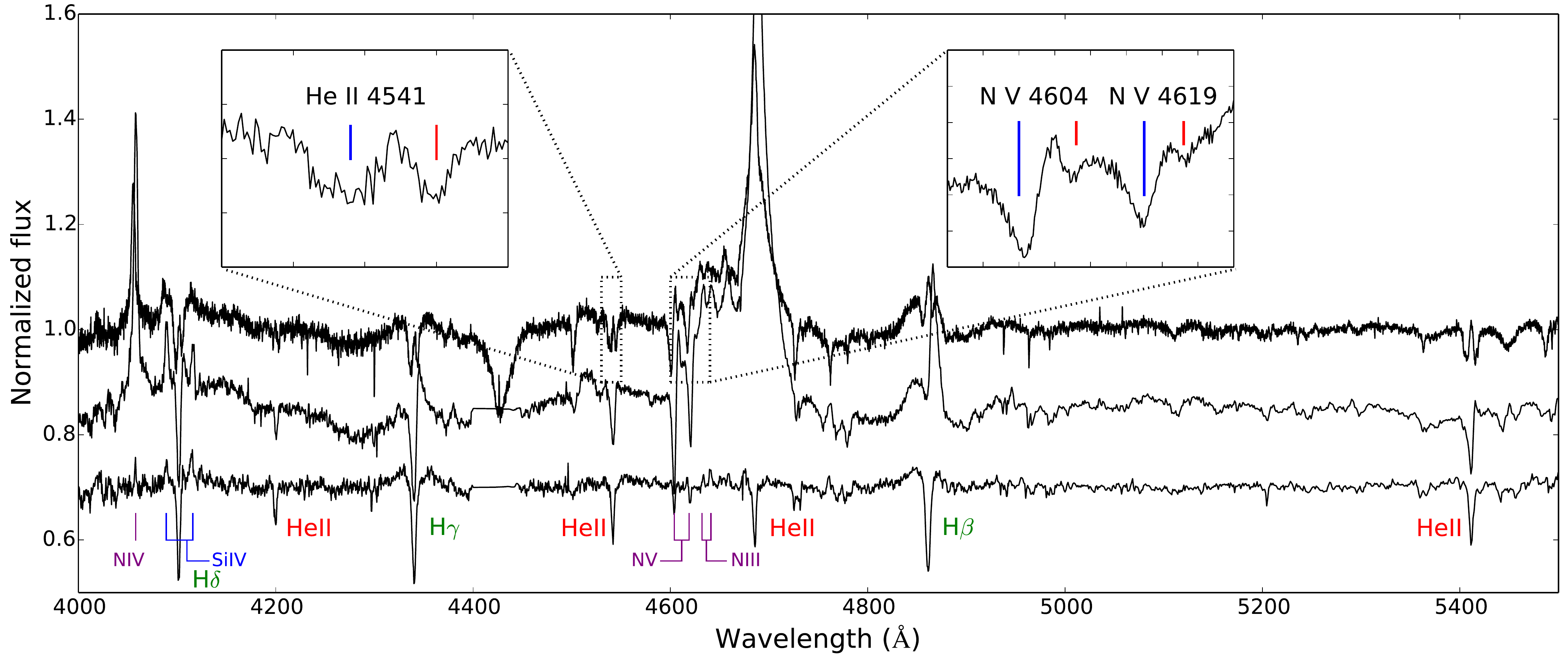}
\caption{Rectified X-Shooter UVB spectrum of WR21a at epoch 16 (top spectrum), and the disentangled spectra of the primary and secondary components (middle and bottom spectra, respectively, and both displaced from unity for clarity). The insets show two of the SB2 lines used for the RV measurements, in which the redshifted contribution of the primary is indicated by the blue line and the blueshifted contribution of the secondary in red.}
\label{fig:spectrum}
\end{figure*}

Massive stars are predominantly found in binary systems, with an intrinsic multiplicity fraction among the O-star population larger than 90\% \citep[e.g., ][]{sana2014}. The orbital properties of massive close binary systems provide a means to address one of the most intriguing open problems in massive star research:  how massive can a star be? If there is an upper mass limit, its value may be dependent on environmental conditions and whether the star forms in solitude or as part of a multiple system. It could be envisioned that the upper mass limit for the components of binary systems is different than for single stars. 

While an upper mass limit of 150 \msun \ for presumed single stars has been derived by \cite{Fig05} based on a photometric analysis of the Arches cluster ($Z\approx 2Z_{\odot}$), stars well in excess of 150 \msun\ have been identified in the core of the R136 region in the Large Magellanic Cloud \citep[$Z\approx 0.5 Z_{\odot}$; the most massive one being R136-a1, with an estimated current mass of 265 \msun;][]{crowther2010}. 

However, the masses of these stars cannot be estimated from their spectral characteristics, as their hydrostatic layers are obscured by their dense stellar winds \citep{dekoter1997}. Instead their masses are inferred from their luminosities, and are model dependent. The luminosities can also be affected by a contribution from unresolved stars if the field is crowded, or by uncertainties in extinction \citep[e.g., ][]{bestenlehner2011}. The orbital motion of binaries in which one or both of the components is a very massive star allows for accurate estimates of the stellar masses with a minimum of model assumptions.  If such systems have not yet experienced interaction, these are most easily linked to the initial masses of the components.

The most massive binary systems in the Galaxy with dynamically measured masses that are currently known are NGC3603-A1 \citep[$116\pm31$ and $89\pm16$ \msun; ][]{schnurr2008}, WR20a \citep[$83\pm5$ and $82\pm5$ \msun; ][]{rauw2004, bonanos2004}. Other very massive binary systems include WR25 \citep{gamen2006} and WR21a. The latter, WR21a, is a possible addition to the list of most massive stars. WR21a is composed of a hydrogen-rich nitrogen sequence Wolf-Rayet (WNha) star and an early-type O star \citep[][henceforth N08]{NGB08}. 

The WNh component of WR21a is the first Wolf-Rayet star detected through its bright X-ray emission \citep{caraveo1989, mereghetti1994}. WR21a also has weak non-thermal radio emission \citep{benaglia2005}. Both the X-ray and radio emission are indicative of a colliding-wind region. Its projected location close to Westerlund 2 suggests that WR21a might have been ejected from this cluster.

N08 performed the first multi-epoch spectroscopic study of WR21a, and found an orbital period $P = 31.673 \pm 0.002$~d and an eccentricity $e = 0.64 \pm 0.03$. They infer minimum masses of 87 \msun \ for the WN star and 53 \msun \ for the O star, placing the system amongst the highest-mass Galactic binaries known. However, the orbital solution of N08 for the secondary component is based on a limited number of measurements of \heb\l 5412. New high-quality data are required to better constrain the orbital and physical properties of this corner-stone system. 

As part of the NOVA program for VLT/X-Shooter guaranteed time, we obtained multi-epoch, intermediate- to high-resolution spectra of WR21a. Our spectra reveal multiple SB2 lines, which allow for a firm determination of the orbit of both components and of their minimum masses and spectral types. In this paper, we use these constraints to estimate the absolute masses of the WR21a components. A quantitative spectroscopic analysis will be the focus of a later work.

This paper is organized as follows. The next Section gives an overview of the observations and data reduction. Section~\ref{sec:orbit} presents the radial-velocity measurements and derives the new orbital solution. Section~\ref{sec:discussion} discusses absolute mass estimates and the systems origin and evolutionary state. We summarize our findings in Section~\ref{sec:conclusions}.

\section{Observations and data reduction}

All observations of WR21a have been obtained with the X-Shooter spectrograph \citep{VDDO11_short} mounted on the European Southern Observatory's {\it Very Large Telescope}. The observational campaign has been designed to intensively cover the January 2011 periastron passage, as well as to provide coverage of the orbit on longer timescales. The journal of the observations including radial-velocity measurements (see Section~\ref{sec:orbit}) can be found in Appendix~\ref{sec:observations}.

The data have been obtained in nodding mode, with a minimum of two exposures per data set. Each individual observation had an exposure time of 90 s, and was obtained with a 0.5\arcsec \ slit for the UVB arm (3000 to 5500 \AA) and  0.4\arcsec\ slits for the VIS  (5500 to 10\,000\AA) and  NIR arms (10\,000 to 25\,000 \AA). These slit widths result in a spectral resolving power $R = \lambda / \Delta \lambda$ of 9900, 18\,200, and 10\,500 in the three arms, respectively. Only the data from the UVB arm are used in the present paper.

The data have been reduced using the X-Shooter pipeline v2.4.0 under the automated {\sc reflex} environment. The resulting 1D flux-calibrated spectra were normalized by fitting a sixth degree polynomial to the continuum and dividing the flux by the resulting function. Finally, the data were corrected for the barycentric motion at the time of observations. The last four epochs in our time series (\#23-26) present low-frequency continuum fluctuations that may be related to  poor performances of the atmospheric dispersion corrector (ADC) and/or poorly defined blaze functions. A local renormalization was applied to compensate for these effects so that they do not impact our radial-velocity (RV) measurements.

\begin{figure}
\centering
\includegraphics[width=\columnwidth]{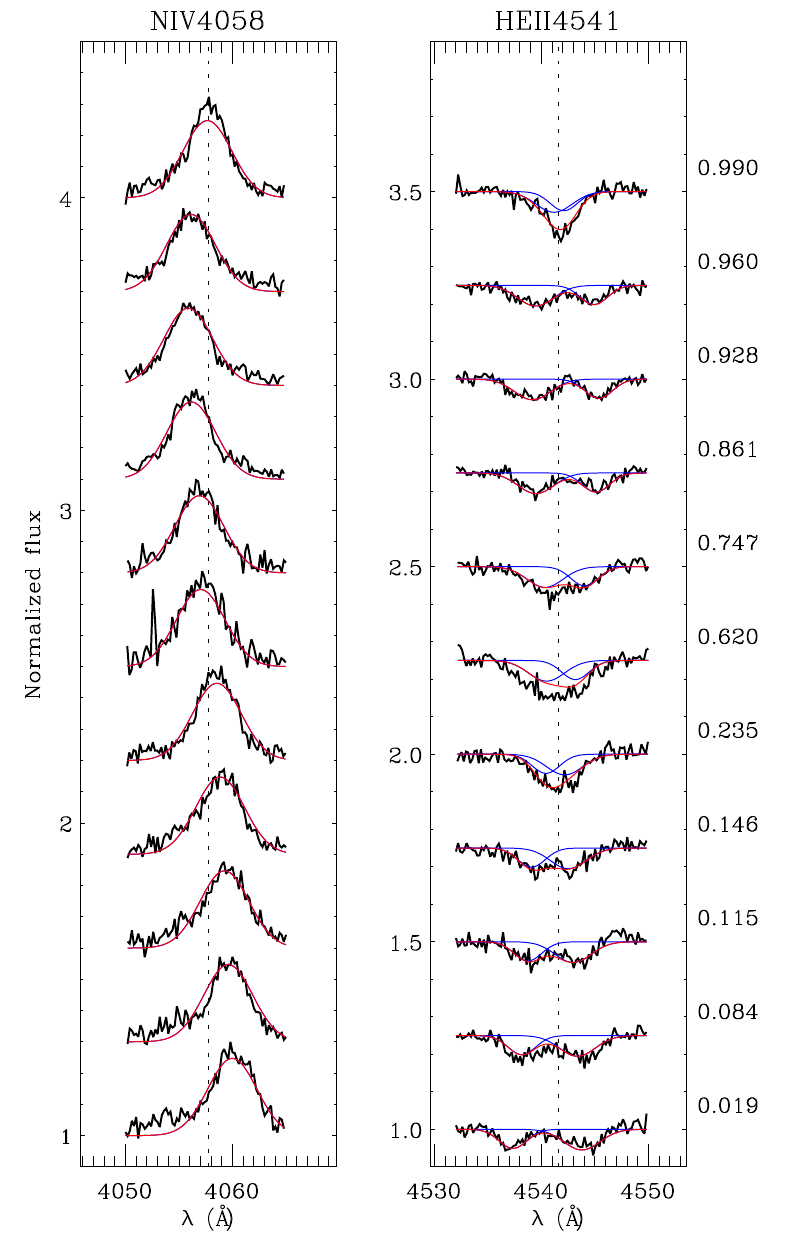}
\caption{Variations of the \nd\l4058 and \heb\l4541 lines as a function of the phase (right-hand side scale). The best-fit Gaussian profiles used for RV determination have been overlaid.}
\label{fig:epochs}
\end{figure}

\section{Orbital properties}\label{sec:orbit}

\subsection{Radial Velocities}
The  X-Shooter UVB  spectra of \wr21a\ present a mix of absorption and emission lines of predominantly H, He and N (see Fig.~\ref{fig:spectrum}). All the lines are variable in position and some of them also in shape. The strongest emission lines are those of \heb\l4686 and \nd\l4058 (see Fig.~\ref{fig:epochs}), which are single-lined and trace the orbital motion of the WR component. \hbet\ presents a  strongly variable  profile with mixed absorption and emission. \hgam, \hdelta\ and \ne\ll4604-4619 are double-lined and dominated by absorption but exhibit a small emission component on their red wing. This indicates that one of the two components may have a weak P-Cygni profile. Finally, the \heb\ll4200, 4541 and 5412 lines show a clear double-lined profile with both components in absorption and with intensities of the order of 5\%\ or less of the continuum level. No trace of \hea\ is seen in the \wr21a\ spectrum, suggesting both stars to be  hot enough to fully ionize helium. High temperatures are also supported by the presence of  \ne\ lines in both stars. 

To measure the radial velocities (RVs) of the two components, we focus on the \heb\ and H8 absorption lines and on the \nd\l4058 emission. We first fitted each spectral line separately using one or two Gaussian functions for single- and double-lined profiles, respectively. We follow the approach described in \citet{SdKdM13} and successfully applied to X-Shooter data in \citet{SvBT13}.  For each spectral line, we fitted one Gaussian profile for each of the two components and we used the same Gaussian profiles to fit all epochs simultaneously. With the exception of two spectra taken at orbital phase $\phi \approx 0.6 - 0.7$ and discussed further below, no strong line profile variations are seen for the considered lines, so that the assumption of keeping the shape of the line profiles fixed is valid. The adopted approach greatly improved the robustness of the RV measurements for our lower signal-to-noise ratio data and for epochs where the lines of the two components are blended.

The RVs measured from the individual lines appear to be compatible with each other, but for a possible systematic shift. Such a shift may be the result of the lines forming in different zones of the outflowing atmospheres. We thus decided to fit all the lines simultaneously. We required that, for a given epoch and a given stellar component, all the lines yield the same RV shift. As additional free parameters, we added a systematic zero point shift ($\delta \gamma$) of the RV measured from each line compared to the systemic velocity ($\gamma$) of the \heb\l4541 line \citep[see, e.g., ][]{taylor2011}. This brings all the measured velocities back onto the \heb\l4541 RV reference frame.

\begin{table*}
\centering
\caption{Best-fit orbital solutions}
\label{tab:os}
\begin{tabular}{llcccc}
\hline
             &       & \multicolumn{2}{c}{All data} & \multicolumn{2}{c}{Cleaned}\\
Parameter    & Units & WR  & O  & WR  & O  \\
\hline
$P$          & [d]          & \multicolumn{2}{c}{$31.680 \pm 0.013$}& \multicolumn{2}{c}{$31.672 \pm 0.011$}  \\
$e$          &              & \multicolumn{2}{c}{$ 0.694 \pm 0.005$}& \multicolumn{2}{c}{$0.6949 \pm 0.0047$} \\
$q$          & $M_1/M_2$    & \multicolumn{2}{c}{$ 1.782 \pm 0.030$}& \multicolumn{2}{c}{$ 1.776 \pm 0.025$} \\
$T$          &$-2\,450\,000$& \multicolumn{2}{c}{$6345.43 \pm 0.32$}& \multicolumn{2}{c}{$ 6345.20 \pm 0.28$} \\
$\omega$     & [\degr]      & \multicolumn{2}{c}{$287.8 \pm 1.2$}   & \multicolumn{2}{c}{$287.2 \pm 1.1$} \\
$K$          & [\kms]       & $157.0\pm 2.3$ & $279.8 \pm 6.2$      & $156.9 \pm 2.0$ & $278.6 \pm 5.3$ \\
$\gamma$     & [\kms]       & $-32.8\pm 1.7$ & $ 32.8 \pm 2.9$      & $-32.5 \pm 1.6$ &  $32.0 \pm 2.7$ \\
$M \sin^3 i$ & [\msun]      & $65.3 \pm 5.6$ & $ 36.6 \pm 1.9$      & $64.4 \pm 4.8$ & $36.3 \pm 1.7$\\
rms          & [\kms]       & \multicolumn{2}{c}{$8.44$}            & \multicolumn{2}{c}{$6.98$} \\
\hline
$\delta \gamma(HeII\lambda4200)$ & [\kms] & \multicolumn{2}{c}{$5.1\pm4.0$} & \multicolumn{2}{c}{$4.8\pm4.1$}\\
$\delta \gamma(HeII\lambda5412$ & [\kms] & \multicolumn{2}{c}{$-6.0\pm2.6$}& \multicolumn{2}{c}{$-5.7\pm2.7$}\\
$\delta \gamma(H8)       $ & [\kms] & \multicolumn{2}{c}{$-5.2\pm2.5$}& \multicolumn{2}{c}{$-5.9\pm2.6$}\\
$\delta \gamma(NIV\lambda4058)  $ & [\kms] & \multicolumn{2}{c}{$19.2\pm2.4$}& \multicolumn{2}{c}{$19.8\pm2.5$}\\
\hline
\end{tabular}
\end{table*}

\subsection{Orbital solution}

\begin{figure*}
\includegraphics[width=6.8cm]{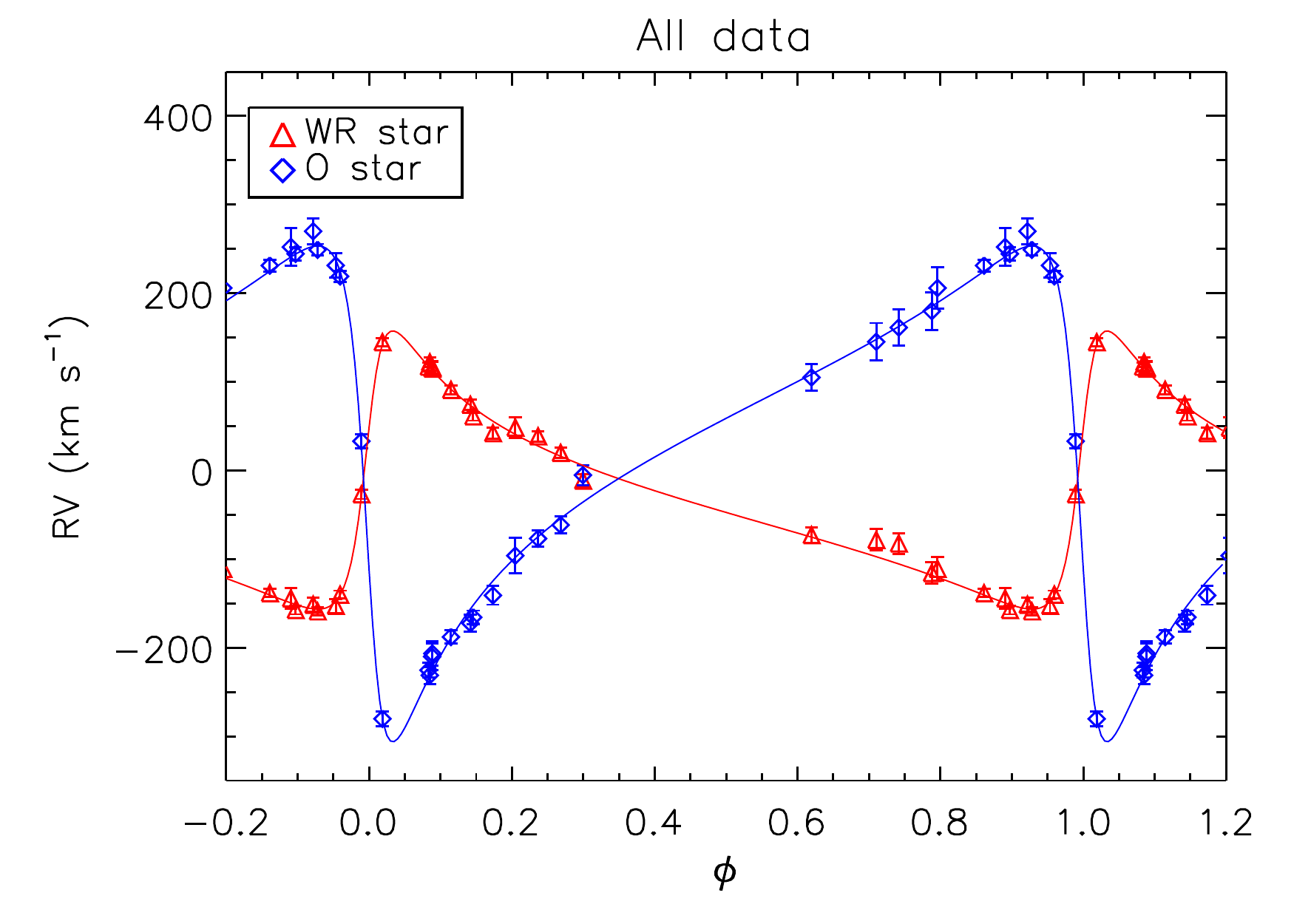}
\includegraphics[width=6.8cm]{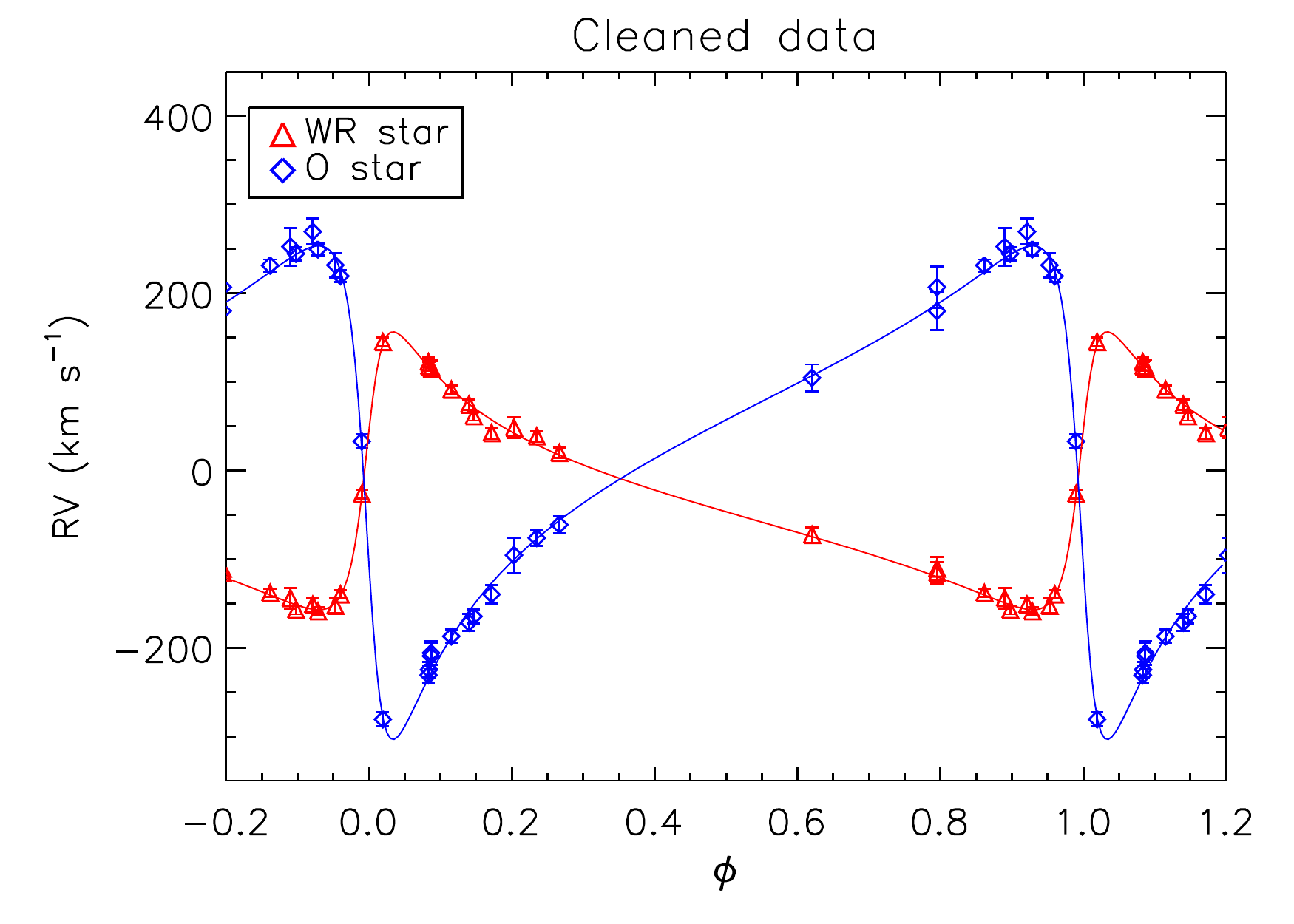}
\caption{Best-fit orbital solutions obtained with all data (left) and with poor RV measurements removed (right).}
\label{fig: os_all}
\end{figure*}

We computed an orbital solution using the RVs obtained by fitting the H8, \nd\l4058 and \heb\ll4200, 4541, 5412 lines simultaneously. In this solution, three epochs provide larger residuals. These are at $\phi=0.3$ (epoch \#7), where the lines are strongly blended, and  at $\phi= 0.72$  and 0.75 (epochs \#24 and 25). For the latter two epochs, \heb\l4541 is stronger than at other phases while H8 seems slightly weaker. The effect is however not as strong as for epoch \#23 at $\phi=0.62$ (Fig.~\ref{fig:epochs}). The change of line intensity seems most significant for the primary component. The \nd\l4058 primary line and the \heb\l5412 line, however, do not show correlated variations. Our final orbital solution excludes these  epochs \#7, 24 and 25 from the fit (see Figure~\ref{fig: os_all}). This results in slightly smaller residuals and error bars, but does not change significantly the fitted parameters (see Table~\ref{tab:os}). 

Although over 50 epochs of observations are available from N08, we chose not to use these for our orbital solution. The selection of spectral lines that we use for our RV measurements are not available in N08 due to the lower signal-to-noise ratio, spectral resolution, and/or wavelength coverage of their data. Instead, N08 had to rely on somewhat challenging lines for their RV measurements (e.g., lines possibly affected by wind-wind collision contamination, by P-Cygni profiles, and/or by severe blending either between the contributions from the primary and the secondary or with nearby interstellar bands). Hence we feared that combining the N08 data set with ours may introduce systemetic biases, so that the resulting solution would not be more reliable than our standalone, homogeneously derived solution.

We find an orbital period of $P=31.672\pm0.011$ d and a highly eccentric orbit ($e = 0.695\pm0.005$), which is similar to the values found by N08 ($P=31.673\pm0.002$ d and $e = 0.64\pm0.03$). As our data show multiple SB2 lines at all epochs and our epochs cover the periastron passage, the orbit of both components is well constrained.

The minimum masses that we find are $M_1 \sin^3 i = 64.4\pm4.8$ \msun \ and $M_2 \sin^3 i = 36.3\pm1.7$ \msun. These are considerably lower than the estimated 87 \msun \ and 53 \msun \ of N08. This results from the slightly larger eccentricity of our orbit compared to that of N08, and a difference in the derived semi-amplitudes of the RV curves.

\section{Discussion}\label{sec:discussion}

In this Section we first derive the spectral types of both components, which give a crude estimate of their masses. We refine these estimates by using the total luminosity of the system as a mass indicator. Finally we discuss the potential origin of WR21a from Westerlund 2 and its evolutionary state.

\subsection{Spectral types}

To determine the spectral types of the components of WR21a, we first apply a new version of the disentangling code presented by \cite{mahy2012} with, as third component, a spectral contribution from the interstellar medium. This code is based on the separation technique developed by \cite{marchenko1998} and refined by \cite{gonzalez2006}, but uses Nelder \& Mead's Downhill Simplex on the orbital parameters to reach the best $\chi^2$ fit between the recombined component spectra and the observed data. In this process, we excluded the four spectra showing low frequency oscillation in the continuum (epochs \#23 to 26), as described earlier. The disentangled spectra are shown in Fig.~\ref{fig:spectrum} and allow for an accurate spectral typing.

The primary late-type WN star shows hydrogen in its spectrum, and is thus of the WNh type. The presence of the \nc \ll4634-41\,\AA \ band and the absence of \nc \l5314\,\AA \ indicates a spectral subtype of WN5, following the \cite{smith1996} system. The primary spectrum shows absorption lines of \heb \ and \ne, and thus the `a' suffix needs to be added \citep{smith1996}. The presence of strong \ne \ absorption indicates that this might be a transition star (`slash' star). We thus derive a spectral type for the primary of O3/WN5ha.

The spectrum of the secondary shows no sign of \hea, indicating a spectral type earlier than O4. This is supported by the presence of \ne \ll4604-4619  absorption lines. The presence of \nc \ emission lines excludes the O2 and O2.5 subtypes \citep{walborn2002}. The relative strengths of the \nc, \nd, and \ne \ lines firmly constrain the spectral subtype to O3. The secondary has a strong \heb \l4686 absorption line, which is deeper than both \heb \l4200 and \heb \l4541, implying a Vz luminosity class. However, \heb \l4686 is likely affected by the colliding wind region and the disentangled profile might not be representative for the stellar spectrum, thus the z tag needs to be taken with caution. Finally, the presence of \nd \l4058, \sid \l4089 and \l4116, and weak emission of \nc \l4634-41 leads us to assign the secondary the spectral type O3Vz~((f*)). 

To provide a first estimation of the absolute masses of the components, we use the \cite{martins2005} spectral type calibration to estimate the mass of the secondary. We do this because the secondary follows a more standard classification scheme than the O3/WN5ha primary star, and thus more reliable mass calibrations exist. The mass of the secondary allows us to estimate the orbital inclination and primary mass from the orbital solution (Table~\ref{tab:os}). 

Given that we can safely exclude the O2~V and O4~V spectral sub-types for the O star (see above), we consider that a one sub-type uncertainty on the spectral type corresponds to a 3$\sigma$ error bar on the mass. Following \cite{martins2005}, we thus assign a typical mass of $58.3\pm3.7$ \msun \ to the O3V secondary with a 1$\sigma$ error bar.  We derive an inclination of $58.8\pm2.5$\degr\ and, correspondingly, a primary mass of $103.6\pm10.2$ \msun.

\subsection{Absolute mass estimates}

To further investigate the absolute masses of the components, we use the total luminosity of the system as a probe for the total mass. To do this, various assumptions have to be made. First, the reddening and total-to-selective extinction in the region of WR21a is highly variable. Here, we use the V-band magnitude and reddening from N08 and \cite{romanlopes2011}. This translates in a reddening corrected apparent V-band magnitude of $m_V = 7.3$. Second, the distance to WR21a is uncertain. Assuming the star originates from Westerlund 2, the distance to this cluster can be used as a probe for the distance to WR21a. However, several distance determinations to Westerlund 2 have been published, with estimates ranging between approximately 2 and 8 kpc. To account for this uncertainty, we calculate the luminosity for three recently published distances: 2.85 kpc \citep{carraro2013}, 6 kpc \citep{HPS15}, and 8 kpc \citep{rauw2007, moffat1991}. Third, a bolometric correction has to be assumed. As the luminosity is dominated by the WNh component, we assume a bolometric correction of 4.0, which is typical for WNh stars \citep{bestenlehner2014}.

\begin{table}
\centering
\caption{Adopted absolute V-band and bolometric magnitudes and their corresponding luminosities for three possible distances.}
\label{tab:magnitudes}
\begin{tabular}{l c c c}
\hline
$d$		& $M_V$	& $M_{\mathrm{bol}}$	& $\log{L/L_{\odot}}$\\
(kpc)\\
\hline
$2.85$		& $-4.96$	& $-8.96$		& $5.48$\\
$6.0$		& $-6.58$	& $-10.58$	& $6.13$\\
$8.0$		& $-7.20$	& $-11.20$	& $6.38$\\
\hline
\end{tabular}
\end{table}

Table~\ref{tab:magnitudes} lists the resulting absolute V and bolometric magnitudes and the corresponding luminosities for each of the above three distances. To translate the luminosities into masses, we use the Baysian interpolation method {\sc bonnsai}\footnote{The {\sc bonnsai} web-service is available at www.astro.uni-bonn.de/stars/bonnsai.} \citep{schneider2014} to compare the luminosities with evolutionary tracks. This provides us with statistically meaningful error bars on the derived masses. This approach assumes that the stars have no prior history of interaction.

Although WR21a is a Galactic star, we use the evolutionary tracks for LMC metallicity \citep{brott2011,kohler2015} to probe the mass-luminosity relation appropriate for the total system mass, as these are the only tracks currently available in {\sc bonnsai} that reach luminosities high enough for our purpose. We use a Salpeter initial mass function \citep{salpeter1955} and the rotational velocity distribution from \cite{ramirez2013} as priors. We derive the mass-luminosity relation for two ages of the system: the components being close to the zero-age main sequence (ZAMS, input age $0.1 \pm 0.1$ Myr) and the components having an age comparable to Westerlund 2 (input age $1.5 \pm 0.5$ Myr). 

\begin{figure}
	\includegraphics[width=\columnwidth]{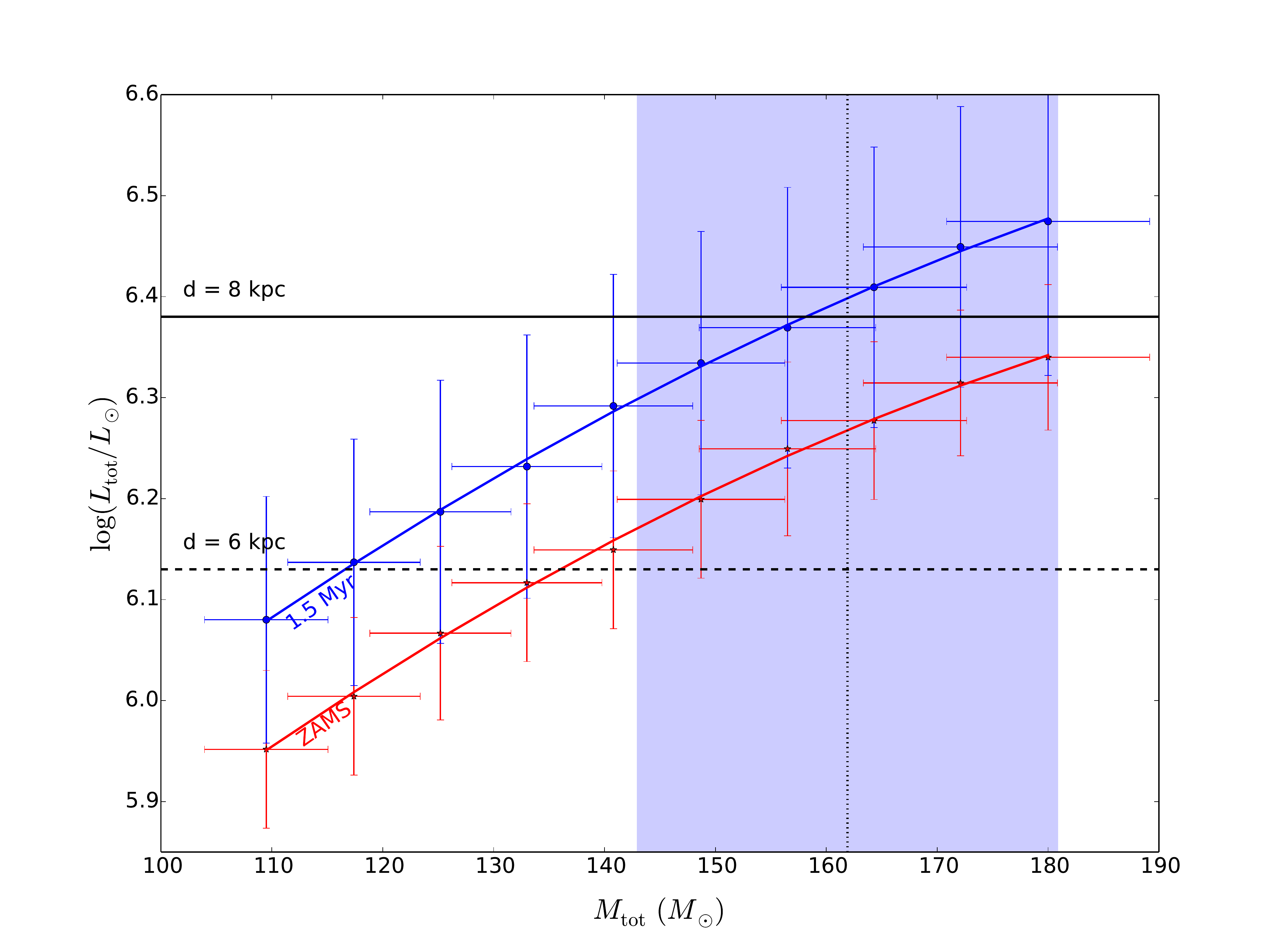}
	\caption{Total mass versus total luminosity relations derived from evolutionary tracks for ages close to the ZAMS (red) and $1.5 \pm 0.5$ Myr (blue). Also indicated are the observed total luminosities for distances of 6 kpc (dashed line) and 8 kpc (solid line). The total system mass that is estimated from the spectral type of the secondary is indicated with the shaded region, and is in agreement with a distance of 8 kpc and an age of 1.5 Myr.}
	\label{fig:ML}
\end{figure}

We derive luminosities for current primary masses of $70 \,M_{\odot} \leq M_1 \leq 115 \, M_{\odot}$ in steps of $5 \,M_{\odot}$, and corresponding secondary masses following from the mass ratio of our orbital solution (Table~\ref{tab:os}). The relative error bars on the masses are set from the orbital solution as well. The resulting total-mass versus total-luminosity relations are shown in Figure~\ref{fig:ML}, and the grid of masses and luminosities is given in Table~\ref{tab:ML}.

Figure~\ref{fig:ML} additionally shows the derived system luminosities for a distance of 6 and 8 kpc from Table~\ref{tab:magnitudes}. Our total mass estimate derived from the spectral type of the secondary is indicated as well. It can immediately be concluded that a distance of 2.85 kpc can be excluded for WR21a, as this corresponds to current masses that are a factor of two lower than the minimum masses obtained from our orbital solution.

Assuming that the distance to WR21a is indeed between 6 and 8 kpc, Figure~\ref{fig:ML} shows that the current system mass is approximately $130 \,M_{\odot} \leq M_{\mathrm{tot}} \leq \,200 M_{\odot}$ if the system is close to the ZAMS, and approximately $110\, M_{\odot} \leq M_{\mathrm{tot}} \leq \, 170 M_{\odot}$ for an age of 1.5 Myr. It is interesting to see that the total system mass of $M_{\mathrm{tot}} = 161.9\pm 19.0 \, M_{\odot}$ that is derived from the spectral type of the secondary is in excellent agreement with the system luminosity at a distance of 8kpc and an age of 1.5 Myr.

The uncertainty in our mass estimates imposed by the use of LMC evolutionary tracks is dominated by the amount of mass that is lost during the evolution. To quantify this difference in mass, we estimate the mass loss at LMC metallicity and scale this to the solar value. To do this, we use our mass estimate of the secondary star as $\it initial$ mass (as we have meaningful error bars on this value). We then use {\sc bonnsai} to quantify the amount of mass that this star would lose in 1.5 Myr at LMC metallicity. This mass loss amounts to $\Delta M_{2, 1.5 \mathrm{Myr}} \approx 2.5 \, M_{\odot}$. Given that the mass loss is proportional to $Z^{0.7}$ \citep[e.g., ][]{vink2001, mokiem2007}, the mass loss of the secondary amounts to $\Delta M_2 \approx 4.1 \, M_{\odot}$ at solar metallicity. In this example, the total current system mass is $155.0 \pm 13.7 \, M_{\odot}$ for LMC metallicity, and $150.4 \pm 13.7 \, M_{\odot}$ for Galactic metallicity. Thus, the uncertainty in the total mass imposed through our use of LMC tracks is on the order of $5 \, M_{\odot}$ for a secondary mass of around $50-60 \, M_{\odot}$, and lower for a lower mass. The derived total masses for LMC and Galactic metallicity remain well within error bars of each other. 

Finally, we compare a total system luminosity of $\log(L/L_{\odot}) = 6.38$ ($\log(L_1/L_{\odot}) \approx 6.25$ and $\log(L_2/L_{\odot}) \approx 5.8$) to the tracks for Galactic metallicity from \cite{ekstrom2012}, which go up to $120\,M_{\odot}$. While this comparison is much less accurate than the {\sc bonnsai} method, these luminosities correspond to masses of approximately $M_1 \approx 100\,M_{\odot}$ and $M_2 \approx 60\, M_{\odot}$ at the ZAMS. This is in excellent agreement with our results from the LMC tracks.

\subsection{Origin and evolutionary state}

Due to its projected distance of 16\arcmin\ from the massive Westerlund 2 cluster, it has been suggested that WR21a originates from this cluster. \cite{romanlopes2011} proposed that WR21a might have been dynamically ejected from the Westerlund 2 core, together with the massive binaries WR20aa and WR20c. These stars are in a similar geometrical configuration in the plane of the sky as AE~Aur, $\mu$ Col, and $\iota$ Ori, for which a binary ejection scenario has been proposed \citep{Bla61}. Disruption of the orbit by dynamical interactions may explain the high eccentricity of the WR21a system, which is higher than any other massive binary with similar period \citep{SdMdK12}. The systematic radial velocity of WR21a that follows from our orbital solution is comparable to that of Westerlund\,2 \citep{RSN11}, and thus there is no indication for a significant runaway motion in the line of sight if the system indeed originates from this cluster. At the maximum distance estimate of 8 kpc, the projected distance of WR21a from Westerlund 2 is 37 pc. WR21a would have to travel at a speed of 36 km~s$^{-1}$ along the plane of the sky in order to reach its current position within 1 Myr. This is well within the maximum velocity at which a massive binary can be ejected \citep{PeS12}.

The derived spectral types of the components of WR21a can be used to estimate their evolutionary state. Hydrogen rich WN stars are very massive main sequence stars that are so luminous that their winds are optically thick \citep{dekoter1997}. They are located close to the ZAMS in the Hertzsprung-Russell diagram (HRD) \citep[e.g., ][]{hamann2006}, and thus the primary O3/WN5ha star is expected to be very young, likely around 1--2 Myr. The O3 secondary star is expected to be of similar age, as stars this massive will evolve towards later spectral types within 2 Myr unless they are rotating rapidly \citep[e.g.][]{brott2011}. The full width at half maximum of \heb \l4541 \AA \ ($3.0206\pm0.0996$ \AA) indicates a rotational velocity of $v_{\mathrm{rot}}\sin{i} \approx 66 $ km s$^{-1}$ following the \cite{ramirez2015} calibration ($v_{\mathrm{rot}} = 77$ km s$^{-1}$ for an inclination $i=58.8\degr$). Thus, we have no indication for such rapid rotation.


\section{Summary and conclusions}\label{sec:conclusions}

We have presented multi-epoch observations of the massive binary system WR21a, including a full coverage of the January 2011 periastron passage. The data allow us to obtain orbital solutions for both components of WR21a. We derive very accurate minimum masses of $M_1 \sin^3 i = 64.4\pm4.8$ \msun \ and $M_2 \sin^3 i = 36.3\pm1.7$ \msun.

We derive spectral types O3/WN5ha and O3Vz~((f*)) for the primary and secondary, respectively. From the spectral type of the secondary we estimate an absolute mass of $M_2 = 58.3\pm3.7$ \msun. This indicates an inclination of $i=58.8\pm2.5$\degr\ and a primary mass of $M_1 = 103.6\pm10.2$ \msun, making the primary the second most massive star with dynamically measured mass in the Galaxy. 

The spectral appearance of WR21a is compatible with an age of 1.5 Myr, similar to that of Westerlund 2. A total system mass of $M_{\mathrm{tot}} = 161.9 \pm 19.0 \,M_{\odot}$, derived from the spectral type of the secondary, and a total system luminosity of $\log{(L_{\mathrm{tot}}/L_{\odot})} = 6.38$ for a distance of 8 kpc, is in good agreement with the mass-luminosity relation predicted by evolutionary models. 

Spectral disentangling will allow us to quantitatively analyse the spectra of both components and obtain their stellar and wind parameters. Comparison of these parameters with evolutionary tracks gives a stronger constraint on the ages of the components, and will give more insight in the evolutionary state of the system. The study of the colliding wind region in X-rays may provide independent constraints on the orbital inclination, and allow us to confirm the absolute masses found in this study. These studies are planned for a future publication on this system. Finally, linear broad-band polarimetry as well as a precision light curve may allow us to obtain independent estimates of the orbital inclination \citep{stlouis1987, lamontagne1996}.

\bibliographystyle{mn2e}
\bibliography{literature_new2}

\appendix

\section{Journal of observations and radial velocity measurements}\label{sec:observations}

Table~\ref{tab:obslog} presents the journal of observations. Given are the Heliocentric Julian Date (HJD), exposure time ($t_{\mathrm{exp}}$), average airmass and spectral signal-to-noise ($S/N$) of each epoch. Additionally, the derived phase ($\phi$) and the radial velocity measurements ($RV$) of both components of WR21a are listed.


\begin{table*}
\centering
\caption{Journal of the observations.}
\label{tab:obslog}
\begin{tabular}{l c c c c c c c}
\hline \hline
Epoch   &   HJD$^a$ &   $t_{\mathrm{exp}}$  &   Average & $S/N^b$ & $\phi$ & RV$_\mathrm{WR}$ & RV$_\mathrm{O}$\\
        &   $-2450000$    & ($s$)   &   airmass & & &  (\kms) & (\kms) \\
\hline
 1     & 5302.6722  & $4\times90$  & 1.36  & 56  &  0.083 &   $122.2\pm 5.5$ &  $-231.1\pm 9.6$  \\
 2     & 5304.4679  & $4\times90$  & 1.32  & 52  &  0.140 &    $74.2\pm 5.8$ &  $-171.7\pm 9.7$  \\
 3     & 5305.4683  & $4\times90$  & 1.31  & 46  &  0.171 &    $42.1\pm 6.3$ &  $-139.8\pm10.5$  \\
 4     & 5306.4652  & $4\times90$  & 1.31  & 23  &  0.203 &    $48.2\pm11.4$ &  $ -95.7\pm20.1$  \\
 5     & 5307.4700  & $4\times90$  & 1.29  & 52  &  0.235 &    $38.5\pm 5.8$ &  $ -76.2\pm 9.4$  \\
 6     & 5308.4729  & $4\times90$  & 1.28  & 54  &  0.266 &    $19.8\pm 5.9$ &  $ -61.2\pm 9.6$  \\
 7$^c$ & 5309.4631  & $4\times90$  & 1.30  & 40   &  0.298 &   $-11.4\pm 4.9$ &  $  -5.3\pm 8.1$  \\
 8     & 5334.4549  & $4\times90$  & 1.20  & 34  &  0.087 &   $115.1\pm 8.1$ &  $-209.5\pm15.1$  \\
 9     & 5334.4626  & $4\times90$  & 1.20  & 38  &  0.087 &   $116.6\pm 7.4$ &  $-206.0\pm13.6$  \\
10     & 5451.9898  & $2\times90$  & 2.64  & 20  &  0.795 &  $-110.8\pm13.1$ &  $ 206.8\pm23.3$  \\
11     & 5454.9112  & $4\times90$  & 2.45  & 22  &  0.890 &  $-144.1\pm11.8$ &  $ 252.6\pm21.3$  \\
12     & 5455.8989  & $4\times90$  & 2.63  & 33  &  0.921 &  $-151.6\pm 8.2$ &  $ 269.5\pm14.6$  \\
13     & 5456.8984  & $4\times90$  & 2.59  & 34  &  0.953 &  $-152.8\pm 8.0$ &  $ 231.7\pm13.8$  \\
14     & 5580.6941  & $8\times90$  & 1.37  & 76  &  0.861 &  $-138.4\pm 4.6$ &  $ 231.4\pm6.7$  \\
15     & 5581.8381  & $8\times90$  & 1.22  & 68  &  0.898 &  $-157.5\pm 4.9$ &  $ 244.6\pm7.4$  \\
16     & 5582.8155  & $8\times90$  & 1.20  & 78  &  0.928 &  $-159.0\pm 4.5$ &  $ 249.5\pm6.6$  \\
17     & 5583.8132  & $8\times90$  & 1.20  & 78  &  0.960 &  $-140.0\pm 4.5$ &  $ 219.3\pm6.6$  \\
18     & 5584.7597  & $8\times90$  & 1.21  & 69  &  0.990 &   $-26.7\pm 5.2$ &  $  32.9\pm8.0$  \\
19     & 5585.6929  & $8\times90$  & 1.33  & 67  &  0.019 &   $145.0\pm 4.9$ &  $-280.9\pm8.1$  \\
20     & 5587.7343  & $8\times90$  & 1.23  & 64  &  0.084 &   $117.3\pm 5.1$ &  $-224.9\pm8.5$  \\
21     & 5588.7303  & $8\times90$  & 1.23  & 70  &  0.115 &    $91.2\pm 4.8$ &  $-187.2\pm7.6$  \\
22     & 5589.7163  & $8\times90$  & 1.25  & 68  &  0.146 &    $61.2\pm 4.9$ &  $-164.7\pm7.7$  \\
23     & 5604.7212  & $6\times90$  & 1.20  & 31  &  0.620 &   $-72.7\pm 8.8$ &  $ 104.5\pm15.1$  \\
24$^c$ & 6114.4800  & $6\times90$  & 1.51  & 22  &  0.715 &  $ -78.4\pm 8.5$ &  $ 145.2\pm15.0$  \\
25$^c$ & 6115.4785  & $6\times90$  & 1.51  & 23  &  0.747 &  $ -82.3\pm 8.3$ &  $ 161.2\pm14.6$  \\
26     & 6338.7075  & $6\times90$  & 1.20  & 22  &  0.795 &  $-115.6\pm12.1$ &  $ 179.9\pm21.3$  \\
\hline
\end{tabular}
\flushleft
Notes: \\
(a) Heliocentric Julian Date at mid-exposure; \\
(b) Signal-to-noise ratio per 0.2 \AA \ bin in the $4250-4300$\AA \ region;  \\
(c) Not used in the final solution.
\end{table*}

\section{Grid of {\sc bonnsai} runs}

The {\sc bonnsai} runs used to derive the total-mass versus total-luminosity relations are listed in Table~\ref{tab:ML}. Columns 1--2 give the input masses, column three the corresponding total system mass, columns 4--5 the derived luminosities for both components for an age close to the ZAMS ($0.1 \pm 0.1$ Myr), column 6 the corresponding system luminosity, columns 7--8 the derived luminosities for both components for an age of $1.5\pm0.5$ Myr, and column 9 the corresponding system luminosity.

\begin{table*}
	\centering
	\caption{Grid of {\sc bonnsai} runs.}
	\label{tab:ML}
	\begin{tabular}{c c c c c c c c c}
		\hline\hline
		& & & \multicolumn{3}{c}{ZAMS} & \multicolumn{3}{c}{1.5 $\pm 0.5$ Myr}\\
		$M_1$ & $M_2$ & $M_{\mathrm{tot}}$ & $\log(L_1/L_{\odot})$ & $\log(L_2/L_{\odot})$ & $\log(L_{\mathrm{tot}}/L_{\odot})$ & $\log(L_1/L_{\odot})$ & $\log(L_2/L_{\odot})$ & $\log(L_{\mathrm{tot}}/L_{\odot})$ \\
		$(M_{\odot})$ & $(M_{\odot})$ & $(M_{\odot})$ & &&&&&\\
		\hline
		$70.0\pm5.3$	&$39.5\pm1.9$	&$109.5\pm5.6$ 	& $5.83\pm0.06$	&$5.34\pm0.05$	&$5.95\pm0.08$	&$5.97\pm0.10$	&$5.43\pm0.07$	&$6.08\pm0.12$\\ 
		$75.0\pm5.6$	&$42.4\pm2.0$	&$117.4\pm6.0$	&$5.88\pm0.06$	&$5.40\pm0.05$	&$6.00\pm0.08$	&$6.02\pm0.10$	&$5.51\pm0.07$	&$6.14\pm0.12$\\ 
		$80.0\pm6.0$	&$45.2\pm2.1$	&$125.2\pm6.4$	&$5.94\pm0.07$	&$5.47\pm0.05$	&$6.07\pm0.09$	&$6.07\pm0.11$	&$5.56\pm0.07$	&$6.19\pm0.13$\\ 
		$85.0\pm6.4$	&$48.0\pm2.3$	&$133.0\pm6.8$	&$5.99\pm0.06$	&$5.52\pm0.05$	&$6.12\pm0.08$	&$6.11\pm0.11$	&$5.62\pm0.07$	&$6.23\pm0.13$\\ 
		$90.0\pm6.8$	&$50.8\pm2.4$	&$140.8\pm7.2$	&$6.02\pm0.06$	&$5.56\pm0.05$	&$6.15\pm0.08$	&$6.17\pm0.11$	&$5.68\pm0.07$	&$6.29\pm0.13$\\ 
		$95.0\pm7.1$	&$53.7\pm2.5$	&$148.7\pm7.6$	&$6.07\pm0.06$	&$5.61\pm0.05$	&$6.20\pm0.08$	&$6.21\pm0.11$	&$5.73\pm0.07$	&$6.33\pm0.13$\\ 
		$100.0\pm7.5$	&$56.5\pm2.7$	&$156.5\pm8.0$	&$6.12\pm0.07$	&$5.66\pm0.05$	&$6.25\pm0.09$	&$6.24\pm0.12$	&$5.78\pm0.07$	&$6.37\pm0.14$\\ 
		$105.0\pm7.9$	&$59.3\pm2.8$	&$164.3\pm8.4$	&$6.14\pm0.06$	&$5.71\pm0.05$	&$6.28\pm0.08$	&$6.28\pm0.12$	&$5.82\pm0.07$	&$6.41\pm0.14$\\ 
		$110.0\pm8.3$	&$62.1\pm2.9$	&$172.1\pm8.8$	&$6.18\pm0.06$	&$5.74\pm0.04$	&$6.31\pm0.07$	&$6.32\pm0.12$	&$5.86\pm0.07$	&$6.45\pm0.14$\\ 
		$115.0\pm8.6$	&$65.0\pm3.1$	&$180.0\pm9.2$	&$6.20\pm0.06$	&$5.78\pm0.04$	&$6.34\pm0.07$	&$6.34\pm0.13$	&$5.90\pm0.08$	&$6.47\pm0.15$\\ 
		\hline
	\end{tabular}
\end{table*}

\label{lastpage}

\end{document}